`

# Influence of the Radial Index on the Stability of Laguerre-Gaussian Vortex Beams in Turbulent Media

A. S. Losev (https://orcid.org/0000-0002-3833-6584)[a,b,c,*], A. Fominova [a],

N. I. Matveeva (https://orcid.org/0000-0002-3580-0994)[a]

[a] *Saint Petersburg State University, Saint-Petersburg, 199034 Russia*
[b] *Saint Petersburg State Marine Technical University, Saint-Petersburg, 190121 Russia*
[c] *ITMO University, Saint-Petersburg, 197101 Russia*

**e-mail: a.losev@spbu.ru*

**Abstract** — This paper explores the selective suppression of Laguerre-Gaussian modes that are most vulnerable to atmospheric turbulence. Decomposing these modes into an orthogonal Zernike polynomial basis reveals significant differences in stability depending on the radial and azimuthal indices. We demonstrate that modes with a higher radial index exhibit minimal distortion of the transverse beam profile, providing a clear criterion for filtering out less resilient modes in turbulent media. Furthermore, we derive an analytical expression relating the required receiver aperture to the radial and azimuthal indices.

## INTRODUCTION

The study of spatially structured light is highly relevant across a broad spectrum of applications, including ultra-high-resolution microscopy [1], optical trapping and micromanipulation [2–6], and information encoding and transmission in both classical [7–11] and quantum [7, 8, 12–15] systems. However, the principal challenge hindering the widespread adoption of free-space optical (FSO) communication is the degradation of transmitted spatial modes due to atmospheric turbulence.
Although our understanding of these phenomena has advanced significantly since the latter half of the 20th century [9, 16–21], current research on the stability of structured light in turbulent environments includes both studies that demonstrate a stable transverse profile [18, 20, 22, 23, 24] and those that report its instability [19, 25].

Laguerre-Gaussian (LG) beams are of particular interest in this context for several reasons. First, they are eigenmodes of free space, meaning they propagate invariantly in a non-turbulent atmosphere

`

and undergo predictable transformations when passing through optical elements such as lenses [25]. Second, as a prominent class of beams carrying orbital angular momentum (OAM) [5], they are exceptionally well-suited for high-capacity information encoding in FSO communication channels [7, 27–29] and free-space quantum key distribution (QKD) systems [30].

To accurately model such communication channels, it is essential to account for wavefront perturbations induced by random fluctuations in the propagation medium [31–33]. Extensive research has been dedicated to examining the behavior of OAM-carrying beams in turbulent media and the impact of turbulence on their key characteristics [23, 34, 35]. Notably, previous studies have demonstrated that utilizing OAM beams can enhance resilience against beam wander and propagation direction fluctuations in turbulent environments [35, 36].

A powerful method for analyzing the wavefront of such beams is decomposition into an orthogonal Zernike polynomial basis [37]. Zernike polynomials are widely used to characterize wavefront distortions caused by optical aberrations [38]. Beyond providing minimizing computational error accumulation, this basis offers several distinct advantages. Crucially, the decomposition coefficients directly correspond to classical optical aberrations, which can be readily measured experimentally using a Shack–Hartmann wavefront sensor [38, 40]. Furthermore, Zernike polynomials are universally used to characterize turbulence-induced wavefront distortions, and like LG modes, they inherently possess circular symmetry [41], making them a mathematically natural choice for this analysis.

In this paper, we demonstrate that certain Laguerre-Gaussian modes exhibit negligible decomposition coefficients in the Zernike polynomial basis. This property renders them significantly more robust against atmospheric turbulence.

The remainder of this paper is organized as follows. Section 2 provides a detailed discussion of the fundamental properties of LG modes and their propagation dynamics in a turbulent atmosphere. Section 3 presents the decomposition of these modes in the Zernike polynomial basis, utilizing this framework to evaluate their propagation stability. Section 4 investigates the specific modes that are most susceptible to beam deformation. Section 5 provides practical guidelines for determining the required receiver aperture as a function of the radial and azimuthal indices. Finally, the Appendix contains supplementary mathematical derivations.

## LAGUERRE-GAUSSIAN MODES

The complex amplitude distribution of an LG beam in the transverse plane, as a function of the propagation distance $z$, is given by [42]:

$$LG_{p,\ell}(\rho,\varphi,z) = \sqrt{\frac{2p!}{w_z^2\,\pi(p+|\ell|)!}}\left(\frac{\rho\sqrt{2}}{w_z}\right)^{|\ell|} L_p^{|\ell|}\left(\frac{2\rho^2}{w_z^2}\right)\exp\left(-\frac{\rho^2}{w_z^2}\left(1-i\frac{z}{z_R}\right)\right)\exp\left(-i\phi_g(z)\right)\exp(i\ell\varphi). \quad (1)$$

Here, $(\rho,\varphi,z)$ denote the cylindrical coordinates; $p \in \mathbb{Z} \geq 0$ and $\ell \in \mathbb{Z}$ are the radial and azimuthal indices, respectively; $z_R$ is the Rayleigh range [34]; $w_z^2 = w_0^2(1 + z^2/z_R^2)$ is the squared beam radius at a propagation distance $z$ from the beam waist (where $w_0 \equiv w(z=0)$ is the waist radius); $\phi_g(z) = (2p + |\ell| + 1)\arctan(z/z_R)$ is the Gouy phase; and $L_p^{|\ell|}$ is the generalized Laguerre polynomial of radial index $p$ and azimuthal index $|\ell|$.

`

The radial ($p$) and azimuthal ($\ell$) indices determine the transverse spatial profile of the beam. Specifically, the index $p$ dictates the number of concentric rings, resulting in $p+1$ bright rings and $p$ dark rings [42]. The azimuthal index $\ell$, also known as the topological charge, represents the number of $2\pi$ phase windings that occur over one full azimuthal rotation around the propagation axis $z$. The sign of l determines the handedness of this helical wavefront. Furthermore, for any non-zero azimuthal index ($\ell \neq 0$), the beam exhibits a phase singularity at its center, meaning the on-axis intensity is strictly zero.

In atmospheric optical communication, Laguerre-Gaussian beams are employed to transmit information encoded in distinct orbital angular momentum states. A common detection method involves using a CCD camera to capture the interference pattern between the incoming LG beam and a reference Gaussian beam tilted at a small angle. This produces a characteristic 'fork' dislocation, the number of prongs (tines) of which depends on the azimuthal index $\ell$. However, this approach has significant limitations. Seminal works [43, 44] demonstrated that when an LG beam with topological charge $\ell$ propagates through atmospheric turbulence, its central phase singularity decays into $|\ell|$ individual unit-charge singularities ($\ell = \pm 1$) arranged in a ring of non-zero radius. Consequently, the interference pattern will exhibit not a single fork corresponding to, for example, the $LG_{0,3}$ mode, but rather three distinct forks characteristic of the $LG_{0,1}$ mode. This phenomenon severely complicates mode-division multiplexing (MDM) aimed at increasing channel capacity, as it becomes difficult or even impossible to unambiguously identify the original OAM state from a single interference pattern.

Conversely, utilizing LG modes that vary in the radial index circumvents this limitation. During propagation through a turbulent medium, the radial ring structure persists and remains discernible in the interference pattern.

It is evident that under moderate to strong turbulence conditions, identifying LG modes based on the azimuthal index $\ell$ becomes highly challenging, often necessitating the use of adaptive optics techniques. Similar conclusions have been drawn in [45–47], which analyzed the crosstalk between overlapping modes with differing OAM propagating in a turbulent medium. Consequently, transmitting information through an atmospheric optical channel using LG modes with different values of $\ell$, without additional wavefront correction systems, is extremely sensitive to both channel quality and transmission distance.

In contrast, employing the radial index p for discrete mode encoding is highly advantageous. In this scenario, the detection process is significantly simplified, reducing to straightforward counting of the number of rings in the interference pattern or the intensity profile of the received beam.

## THE RELATIONSHIP BETWEEN LAGUERRE-GAUSSIAN MODES AND ZERNIKE POLYNOMIALS

Consider a complete set of Zernike polynomials orthogonal over a circle of radius $w_0$ [41]:

$$Z_n^m(\rho,\varphi) = \sqrt{\frac{n+1}{\pi w_0^2}} R_n^m(\rho) \mathrm{e}^{im\varphi}, \quad (2)$$

where *n* is the radial degree and *m* is the azimuthal frequency. Here,

$$R_n^m(\rho) = \sum_{t=0}^{A_n^{|m|}} (-1)^t \frac{(n-t)!}{t!\left(A_n^{|m|}-t\right)!\left(B_n^{|m|}-t\right)!} \left(\frac{\rho}{w_0}\right)^{n-2t} \quad (3)$$

`

are the radial polynomials, with $A_n^{|m|} = (n - |m|)/2$ and $B_n^{|m|} = (n + |m|)/2$.

Each Zernike polynomial describes a specific type and order of optical aberration. When an arbitrary wavefront is decomposed into a Zernike series, the expansion coefficients quantify the contribution of each aberration type. Aberrations play a critical role in the propagation of spatially structured beams. For instance, the disintegration of the phase singularity at the center of an LG beam can be attributed to initial astigmatism [48], which may result in dynamic OAM inversion during propagation through a medium [49]. In Ref. [26], the behavior of beams with wavefronts represented as superpositions of Zernike polynomials was investigated. Consequently, decomposing an arbitrary LG mode into a Zernike polynomial basis allows estimation of the potential distortion experienced by the beam propagating through a turbulent medium. This estimation depends on the values of the radial index $p$ and azimuthal index $\ell$.

We consider the case of a weakly converging beam at distances along the z-axis where the Rayleigh range satisfies $z_R \gg z$. This allows us to simplify expression (1) by setting the ratio $z/z_R$ to zero, effectively considering the transverse beam profile expansion in the waist region at $z = 0$. Using the substitution

$$x = \frac{2\rho^2}{w_0^2} \quad (4)$$

expression (1) takes the form

$$LG_{p,\ell}(x, \varphi) = \frac{1}{w_0}\sqrt{\frac{2p!}{\pi(p+|\ell|)!}}\, x^{|\ell|/2}\, L_p^{|\ell|}(x)\, \mathrm{e}^{-x/2}\, \mathrm{e}^{i\ell\varphi}. \quad (5)$$

The decomposition over the complete set of orthogonal Zernike polynomials can be expressed as

$$LG_{p,\ell}(x, \varphi) = \sum_{n=0}^{\infty} \sum_{m=-n}^{n} S_{p,\ell}^{n,m}\, Z_n^m(x, \varphi), \quad (6)$$

where the expansion coefficients are given by

$$S_{p,\ell}^{n,m} = \frac{w_0^2}{4} \int_0^{2\pi} d\varphi \int_0^2 dx\, LG_{p,\ell}(x, \varphi)\, (Z_n^m(x, \varphi))^*. \quad (7)$$

After integrating over the angular variable and applying the Rodrigues formula for Laguerre polynomials in binomial form [50], we obtain

$$S_{p,\ell}^{n,m} = \sqrt{\frac{(n+1)p!}{2(p+|\ell|)!}} \sum_{s=0}^{p} \sum_{t=0}^{A_n^{|m|}} \frac{(-1)^{s+t}\, 2^{t-\frac{n}{2}}}{s!}\, C_{n-t}^{A_n^{|m|}}\, C_{A_n^{|m|}}^{t}\, C_{p+|\ell|}^{p-s} \int_0^2 dx\, x^{B_n^{|\ell|}+s-t}\, \mathrm{e}^{-\frac{x}{2}}\, \delta_{m,\ell}. \quad (8)$$

The integral over dx is well-known [51, 52]:

$$\int_0^u dx\, x^{\nu-1}\, \mathrm{e}^{-x} = \gamma(\nu, u) = (\nu - 1)!\left(1 - e^{-u} \sum_{k=0}^{\nu-1} \frac{u^k}{k!}\right), \quad Re(\nu) > 0, \quad (9)$$

where $\gamma$ is the lower incomplete gamma function. This yields the final expression for the expansion coefficients:

$$S_{p,\ell}^{n} = \sqrt{\frac{n+1}{2^{n+1}|\ell|!\, C_{p+|\ell|}^{|\ell|}}} \sum_{s=0}^{p} \sum_{t=0}^{A_n^{|\ell|}} \frac{(-1)^{s+t}\, 2^t}{s!}\, C_{A_n^{|\ell|}}^{t}\, C_{n-t}^{A_n^{|\ell|}}\, C_{p+|\ell|}^{p-s}\, \left(B_n^{|\ell|} + s - t\right)! \left(1 - \frac{1}{e} \sum_{k=0}^{B_n^{|\ell|}+s-t} \frac{1}{k!}\right). \quad (10)$$

In this case, the expression (6) takes the form

$$LG_{p,\ell}(x, \varphi) = \sum_{n=0}^{\infty} S_{p,\ell}^{n}\, Z_n^{\ell}(x, \varphi). \quad (11)$$

## THE MODE OF EQUALITY OF RADIAL DEGREE AND AZIMUTHAL INDEX

In Ref. [26], it was shown that Zernike modes with matching radial and azimuthal indices ($n = |\ell|$) undergo the maximum transformation (exhibit the highest root-mean-square error of expansion coefficients) during free-space propagation. This inherent structural instability suggests that

`

minimizing the contribution of such Zernike modes can significantly enhance the overall robustness of the beam in turbulent media.Consequently, minimizing these expansion coefficients will enhance the stability of the corresponding Laguerre-Gaussian modes.

Let us simplify expression (10) for the case $n = |\ell|$:

$$S_{p,\ell}^{(n=|\ell|)} = \frac{1}{\mathrm{e}}\sqrt{\frac{(|\ell|+1)(|\ell|+p)!}{2^{|\ell|+1}\,p!}}\sum_{s=0}^{p}(-1)^s\, C_p^s\, \sum_{k=|\ell|+s+1}^{\infty} \frac{1}{k!}. \quad (12)$$

Using the ascending finite difference operator $\Delta^p f(0) = (-1)^p \sum_{s=0}^{p}(-1)^s\, C_p^s\, f_s$ (with step 1 at coordinate 0) we obtain:

$$S_{p,\ell}^{(n=|\ell|)} = \mathcal{N}_{p,\ell}\,(-1)^p\,\Delta^p f(0), \quad (13)$$

where

$$\mathcal{N}_{p,\ell} = \sqrt{\frac{(|\ell|+1)(|\ell|+p)!}{2^{|\ell|+1}\,p!}} \quad (14)$$

and

$$f_s = \frac{1}{\mathrm{e}}\sum_{k=|\ell|+s+1}^{\infty} \frac{1}{k!}. \quad (15)$$

Although equation (13) is not strictly zero for any $p, |\ell| \in \mathbb{Z} > 0$, it can effectively vanish. As shown in the Appendix, the finite difference converges to zero faster than a geometric progression

$$|\Delta^p f(0)| = O\left(\frac{1}{(|\ell|+p)!}\right). \quad (16)$$

This yields values $S_{p,\ell}^{(n=|\ell|)} \leq 0.04$ for any fixed $|\ell| > 0$ when $p > 8$ (see Table 1). Figure 1 illustrates this behavior: for any given values of $n = |\ell|$, the coefficients exhibit oscillatory dependence on $p$, while the oscillation amplitudes show a general decreasing trend with both increasing $p$ and $|\ell|$.

We note that this decreasing trend holds not only for the specific modes with $n = |\ell|$ under consideration but also for the general expansion coefficients given by equation (10).

Thus, by increasing the radial index $p$ and/or the azimuthal index $|\ell|$, the contribution of unstable Zernike modes with $n = |\ell|$ can be minimized, thereby enhancing the propagation stability of LG beams.

## PRACTICAL GUIDELINES FOR RECEIVER APERTURE

An increase in the radial index p necessarily leads to an expansion of the beam diameter, which may be constrained by the aperture of the detecting device. Here, we estimate the required receiver aperture for $p > 8$. For LG modes with a non-zero radial index p, the intensity is distributed across concentric rings rather than being concentrated within 2–3 effective beam radii. Consequently, estimating the required aperture for <1% energy loss using the second moment of intensity (as defined by the ISO 11146 standard) yields inaccurate results.

Instead, the condition for capturing 99% of the beam's energy within a circle of radius R, using substitution (4), is given by:

$$\frac{\int_0^R d\rho\,\rho\, I_{p,\ell}(\rho)}{\int_0^\infty d\rho\,\rho\, I_{p,\ell}(\rho)} = \frac{\int_0^{2R^2/w_0^2} dx\,|LG_{p,\ell}(x,\varphi)|^2}{\int_0^\infty dx\,|LG_{p,\ell}(x,\varphi)|^2} = \int_0^{2R^2/w_0^2} dx\; x^{|\ell|}\left(L_p^{|\ell|}(x)\right)^2 \mathrm{e}^{-x} = 0.99. \quad (17)$$

Numerical evaluation of this integral for $p$ ranging from 9 to 15 yields a simple approximate formula:

`

$$R_{99} \approx w_0\sqrt{2p+|\ell|+3}. \quad (18)$$

The effective radius of an LG mode is defined as

$$w_{eff} = w_0\sqrt{2p+|\ell|+1}. \quad (19)$$

Therefore,

$$R_{99} \approx w_{eff}\sqrt{1+\frac{2}{2p+|\ell|+1}}. \quad (20)$$

For large $p$, the fraction under the square root tends to zero, yielding $R_{99} \approx w_{eff}$. This implies that 99% of the energy of a high-order LG beam is concentrated within a circle whose radius is approximately equal to its effective radius. This stands in stark contrast to the fundamental Gaussian mode, where only 86.5% of the energy is contained within $w_{eff}$.

Thus, to detect an LG beam with $p > 8$ with negligible energy loss (<1%), the minimum diameter of the receiver aperture (e.g., a CCD or sCMOS camera sensor) must be:

$$D_{aperture} = 2R_{99} \approx 2w_0\sqrt{2p+|\ell|+3}. \quad (21)$$

For typical values of p in this range, this gives approximately $D_{aperture} \sim 10\, w_0$. Beam waist sizes in the range of 0.5–3.0 mm are standard and easily achievable in free-space optical setups using conventional lenses. Furthermore, the typical active area diagonal of modern scientific cameras ranges from 15 to 25 mm. This is more than sufficient to detect modes with $p \leq 15$, provided the beam scale is appropriately matched ($w_0 \approx 1-2$ mm) using a standard lens system.

Thus, while an increase in the radial index $p$ expands the beam diameter and could potentially be constrained by the active area of the detector, this does not represent a fundamental limitation.

## CONCLUSIONS

This study has demonstrated that LG modes with different radial (p) and azimuthal (ℓ) indices exhibit markedly different resilience to propagation through turbulent media. Unlike modes encoded with azimuthal indices, which suffer from the decay of the central fork dislocation into multiple separate singularities, modes utilizing radial indices maintain their concentric ring structure after propagation through turbulence. This structural stability provides a significant advantage for information encoding applications.

A general decreasing trend is observed in the absolute values of the expansion coefficients $\left|S^n_{p,\ell}\right|$ with increasing radial index p, including coefficients for Zernike polynomials with matching indices (n = |ℓ|). Consequently, to enhance noise stability in turbulent channels, it is advisable to employ LG modes with larger radial indices.

For the practically relevant range p = 9 to 15, which provides substantial resilience to atmospheric turbulence, the required detector aperture can be calculated using the simple analytical formula (21). These theoretical predictions are directly applicable to experimental implementations.

## ACKNOWLEDGMENTS

The authors thank D.D. Reshetnikov for fruitful discussions.

`

## CONFLICT OF INTEREST

The authors declare that they have no conflicts of interest.

## APPENDIX

As stated in equation (15) $f_s = \frac{1}{\mathrm{e}} \sum_{k=|\ell|+s+1}^{\infty} \frac{1}{k!}$. Let us derive an upper bound for $f_s$. Note that for $k \geq |\ell| + s + 1$,

$$\frac{1}{k!} \leq \frac{1}{(|\ell|+s+1)!} \cdot \frac{1}{(|\ell|+s+2)^{k-|\ell|-s-1}}. \quad (18)$$

Hence,

$$f_s \leq \frac{1}{\mathrm{e}\,(|\ell|+s+1)!} \sum_{j=0}^{\infty} \frac{1}{(|\ell|+s+2)^j} \quad (19)$$

The sum represents an infinite geometric progression $\sum_{j=0}^{\infty} aq^j = a/(1-q)$ with the first term $a = 1$ and ratio $q = 1/(|\ell| + s + 2)$:

$$\sum_{j=0}^{\infty} \frac{1}{(|\ell|+s+2)^j} = 1 + \frac{1}{|\ell|+s+1} \leq 2. \quad (20)$$

Thus, we obtain the upper bound for $f_s$:

$$f_s \leq \frac{2}{\mathrm{e}\,(|\ell|+s+1)!} \quad (21)$$

Since $f_s$ decreases faster than a geometric progression, we can write

$$f_s \leq C\, q^s \quad (22)$$

where C is a constant and $0 < q < 1$.

To estimate the finite difference, we use the following lemma: if $f_s \leq C\, q^s$, then $|\Delta^p f(0)| \leq C(1+q)^p$. In our case, from equation (21) $f_s \sim \frac{1}{(|\ell|+s)!}$, therefore

$$|\Delta^p f(0)| \leq \frac{1}{(|\ell|+p)!}. \quad (23)$$

`

REFERENCES

1. S. W. Hell and J. Wichmann, Opt. Lett. **19**, 780 (1994). https://doi.org/10.1364/ol.19.000780
2. M. P. MacDonald, L. Paterson, K. Volke-Sepulveda, et al., Science **296**, 1101 (2002). https://doi.org/10.1126/science.1069571
3. A. Jesacher, S. Fürhapter, S. Bernet, et al., Opt. Express. **12**, 4129 (2004). https://doi.org/10.1364/OPEX.12.004129
4. E. G. Abramochkin, K. N. Afanasiev, V. G. Volostnikov et al., Bull. Russ. Acad. Sci. Phys. **72**, 68 (2008). https://doi.org/10.3103/S1062873808010176
5. L. Allen, M. W. Beijersbergen, R. J. C. Spreeuw, J. P. Woerdman, Phys Rev A **45**, 8185 (1992). https://doi.org/10.1103/PhysRevA.45.8185
6. V. V. Kotlyar, E. G. Abramochkin, A. A. Kovalev, et al., Computer Optics 2024; **48**, 180 (2024). https://doi.org/10.18287/2412-6179-CO-1374
7. A. E. Willner, K. Pang, H. Song, et al., Appl. Phys. Rev. **8**, 041312 (2021). https://doi.org/10.1063/5.0054885
8. A. Trichili, K.-H. Park, M. Zghal, et al., IEEE Commun. Surv. Tutorials **21**, 3175 (2019). https://doi.org/10.1109/COMST.2019.2915981
9. M. P. J. Lavery, C. Peuntinger, K. Günthner, et al., Science Advances **3**, 1700552 (2017). https://doi.org/10.1126/sciadv.1700552
10. T. Doster and A. T. Watnik, Applied Optics **55**, 10239 (2016). https://doi.org/10.1364/AO.55.010239
11. A. Trichili, A. Salem, A. Dudley, et al., Opt. Lett. **41**, 3086 (2016). https://doi.org/10.1364/OL.41.003086
12. M. Mirhosseini, O. Magaña-Loaiza, M. O'Sullivan, et al., New J. Phys. **17**, 033033 (2015). https://doi.org/10.1088/1367-2630/17/3/033033
13. Z. Wang, R. Malaney, and J. Green, in 2019 IEEE Global Communications Conference (GLOBECOM), 1–6 (2019). https://doi.org/10.1109/GLOBECOM38437.2019.9014321
14. D. A. Turaykhanov, D. O. Akatev, I. Z. Latypov, et al., Bull. Russ. Acad. Sci. Phys. **84**, 304 (2020). https://doi.org/10.3103/S1062873820030247
15. V. P. Lukin, I. P. Lukin, Computer Optics **48**, 68 (2024). https://doi.org/10.18287/2412-6179-CO-1355.
16. V. I. Tatarski, Science **134**, 3475 (1961). https://doi.org/10.1126/science.134.3475.324.c
17. L. C. Andrews, Journal of Modern Optics **39**, 1849 (1992). https://doi.org/10.1080/09500349214551931
18. W. Cheng, J. W. Haus, and Q. Zhan, Opt. Express **17**, 17829 (2009). https://doi.org/10.1364/OE.17.017829
19. M. A. Cox, C. Rosales-Guzmán, M. P. J. Lavery, et al., Opt. Express **24**, 18105 (2016). https://doi.org/10.1364/OE.24.018105
20. P. Lochab, P. Senthilkumaran, and K. Khare, Phys. Rev. A **98**, 023831 (2018). https://doi.org/10.1103/PhysRevA.98.023831

`

21. M. A. Cox, N. Mphuthi, I. Nape, et al., IEEE J. Sel. Top. Quantum Electron. **27**, 7500521 (2020). https://doi.org/10.1109/JSTQE.2020.3023790
22. O. K. Y. Gu and G. Gbur, Opt. Lett. **34**, 2261 (2009). https://doi.org/10.1364/OL.34.002261
23. A. Klug, C. Peters, and A. Forbes, Adv. Photon. **5**, 016006 (2023). https://doi.org/10.1117/1.AP.5.1.016006
24. A. V. Volyar, E. G. Abramochkin, M. V. Bretsko et al., Journal of Optics **10**, 727 (2023). https://doi.org/10.3390/photonics10070727
25. B. Ndagano, N. Mphuthi, G. Milione, et al., Opt. Lett. **42**, 4175 (2017). https://doi.org/10.1364/OL.42.004175
26. P. Khorin, Journal of Physics: Conference Series **1096**, 012104 (2018). https://doi.org/10.1088/1742-6596/1096/1/012104
27. G. Gibson, M. Graham, J. Courtial, et al., Optics express **12**, 5448 (2004). https://doi.org/10.1364/OPEX.12.005448
28. J. Wang, J. Yang, I. Fazal, et al., Nat. Photonics **6**, 488 (2012). https://doi.org/10.1038/nphoton.2012.138
29. A. E. Willner, H. Huang, Y. Yan, et al., Adv. Opt. Photonics **7**, 66 (2015). https://doi.org/10.1364/AOP.7.000066
30. D. D. Reshetnikov, A. L. Sokolov, E. A. Vashukevich, et al., Radiophys Quantum El. **67**, 51 (2024). https://doi.org/10.1007/s11141-025-10352-z
31. R. Barros, S. Keary, L. Yatcheva, et al., in Remote Sensing of Clouds and the Atmosphere XIX; and Optics in Atmospheric Propagation and Adaptive Systems XVII, A. Comerón, E. I. Kassianov, K. Schäfer, et al., Eds., Proc. SPIE 9242, 92421L (2014). https://doi.org/10.1117/12.2070694
32. V. P. Lukin, Phys. Usp. **10**, 280 (2021). https://doi.org/10.3367/UFNr.2020.10.038849
33. V. Lukin, P. Konyaev, and V. Sennikov, Applied Optics **51**, 176 (2012). https://doi.org/10.1364/AO.51.000C84
34. V. P. Aksenov, V. V. Dudorov, and V. V. Kolosov, Quantum Electron. **46**, 729 (2016). https://doi.org/10.1070/QEL16088
35. V. P. Aksenov and C. E. Pogutsa, Applied optics **51**, 7262 (2012). https://doi.org/10.1364/AO.51.007262
36. D. D. Reshetnikov, T. K. Korol, E. V. Malyutina, et al., Optics and Spectroscopy **132**, 705 (2024). https://doi.org/10.61011/EOS.2024.07.59649.6774-24
37. M. S. Kirilenko, P. A. Khorin, and A. P. Porfirev, in CEUR Workshop Proceedings, 1638, 66 (2016). https://doi.org/10.18287/1613-0073-2016-1638-66-75
38. A. Wada, T. Ohtani, Y. Miyamoto, et al., Journal of the Optical Society of America A. **22**, 2746 (2005). https://doi.org/10.1364/JOSAA.22.002746
39. B. C. Platt, R. Shack, Journal of Refractive Surgery **17**, S573 (2001). https://doi.org/10.3928/1081-597X-20010901-13
40. Q. Yu, Y. Li, M. Qin, X. Hao, Acta Optica Sinica, **45**, 2100001 (2025). https://doi.org/10.3788/AOS251187

`

41. A. Wünsche, Generalized Zernike or disc polynomials **174**, 135 (2005). https://doi.org/10.1016/j.cam.2004.04.004
42. S. Solimeno, B. Crosignani, and P. D. Porto, *Guiding, Diffraction, and Confinement of Optical Radiation* (Elsevier Science, The University of California, 1986).
43. I. V. Basistiy, M. S. Soskin, and M. V. Vasnetsov, Optics Communications **119**, 604 (1995). https://doi.org/10.1016/0030-4018(95)00267-C
44. F. Ricci, W. Lo¨ffler, and M. P. van Exter, Optics Express **20**, 22961 (2012). https://doi.org/10.1364/OE.20.022961
45. S. Zhao, L. Wang, L. Zou, et al., Optics Communications **376**, 92 (2016). https://doi.org/10.1016/j.optcom.2016.04.075
46. L. Li, R. Zhang, Z. Zhao, et al., Sci. Rep. **7**, 17427 (2017). https://doi.org/10.1038/s41598-017-17580-y
47. L. Torner, J. P. Torres, and S. Carrasco, Optics Express. **13**, 873 (2005). https://doi.org/10.1364/OPEX.13.000873
48. I. V. Basistiy, V. Y. Bazhenov, M. S. Soskin, et al., Opt. Commun. 103, 422 (1993). https://doi.org/10.1016/0030-4018(93)90168-5
49. G. Molina-Terriza, J. Recolons, J. P. Torres, et al., Phys. Rev. Lett. **87**, 023902 (2001). https://doi.org/10.1103/PhysRevLett.87.023902
50. T. Koornwinder, R. Wong, R. Koekoek, et al., *NIST handbook of mathematical functions* (Cambridge University Press, Cambridge, 2010).
51. I. S. Gradshteyn and I. M. Ryzhik, *Table of integrals, series, and products* (Translated from the Russian. Translation edited and with a preface by Alan Jeffrey and Daniel Zwillinger, Elsevier/Academic Press, Amsterdam, 2007).
52. G. Arfken, The incomplete gamma function and related functions (3rd ed., Academic Press, Orlando, FL, 1985).

`

# TABLES

**Table 1.** Zernike expansion coefficients $S_{p,\ell}^{(n=|\ell|)}$ as a function of radial index $p$ and azimuthal index $\ell$.

| $S_{p,\ell}^{(n=|\ell|)}$ | $|\ell|$=0 | $|\ell|$=1 | $|\ell|$=2 | $|\ell|$=3 | $|\ell|$=4 | $|\ell|$=5 | $|\ell|$=6 | $|\ell|$=7 | $|\ell|$=8 | $|\ell|$=9 |
|---|---|---|---|---|---|---|---|---|---|---|
| $p$ = 0 | 0.447 | 0.187 | 0.070 | 0.023 | 0.007 | 0.002 | 0.001 | 0.000 | 0.000 | 0.000 |
| $p$ = 1 | 0.260 | 0.184 | 0.092 | 0.038 | 0.013 | 0.004 | 0.001 | 0.000 | 0.000 | 0.000 |
| $p$ = 2 | 0.130 | 0.150 | 0.098 | 0.047 | 0.019 | 0.007 | 0.002 | 0.001 | 0.000 | 0.000 |
| $p$ = 4 | -0.011 | 0.068 | 0.081 | 0.055 | 0.028 | 0.012 | 0.004 | 0.001 | 0.000 | 0.000 |
| $p$ = 5 | -0.041 | 0.033 | 0.065 | 0.054 | 0.031 | 0.014 | 0.006 | 0.002 | 0.001 | 0.000 |
| $p$ = 8 | -0.051 | -0.030 | 0.017 | 0.037 | 0.032 | 0.019 | 0.009 | 0.004 | 0.001 | 0.000 |
| $p$ = 9 | -0.041 | -0.038 | 0.003 | 0.029 | 0.030 | 0.020 | 0.010 | 0.005 | 0.002 | 0.000 |
| $p$ = 10 | -0.028 | -0.041 | -0.008 | 0.021 | 0.027 | 0.020 | 0.011 | 0.005 | 0.002 | 0.001 |
| $p$ = 11 | -0.016 | -0.041 | -0.016 | 0.014 | 0.024 | 0.020 | 0.012 | 0.006 | 0.002 | 0.001 |
| $p$ = 15 | 0.021 | -0.017 | -0.028 | -0.010 | 0.009 | 0.015 | 0.012 | 0.008 | 0.004 | 0.002 |

`

FIGURE CAPTIONS

**Fig. 1.** Zernike expansion coefficients $S_{p,\ell}^{(n=|\ell|)}$ as a function of radial index $p$ for the case $n = |\ell|$.

`

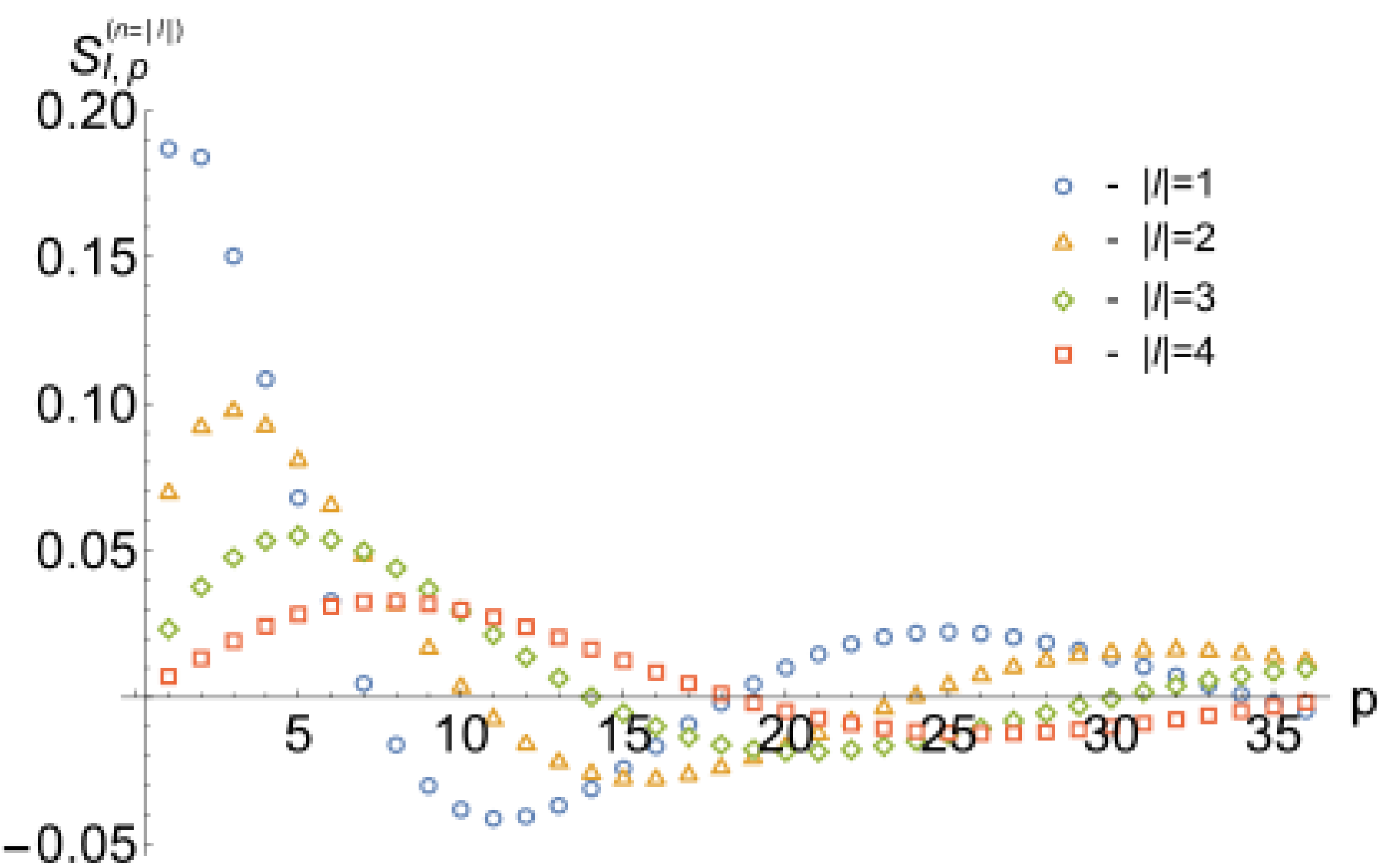

Fig. 1.